\documentclass[aps,prl,twocolumn,superscriptaddress]{revtex4}
\usepackage{natbib}
\usepackage{graphicx}
\usepackage{dcolumn}
\usepackage{amsmath,amsthm}
\usepackage{amsfonts}
\usepackage{amssymb}
\usepackage{bbold}
\usepackage{times}
\usepackage{mathrsfs}
\usepackage{color}
\usepackage{gensymb}
\usepackage{times}
\usepackage{bm}
\usepackage{stmaryrd}
\DeclareGraphicsExtensions{.PNG,.jpg,.pdf,.gif,.eps}
\usepackage{amsmath,amssymb,amsthm,amsfonts,eucal}
\usepackage[english]{babel}
\usepackage{mathtools, nccmath}
\graphicspath{{Figures/}}

\begin{document}


\title{Frictionless motion of lattice defects}

\author{N.Gorbushin}
\affiliation{\it  PMMH, CNRS -- UMR 7636, CNRS, ESPCI Paris, PSL Research University, 10 rue Vauquelin, 75005 Paris, France}

\author{G. Mishuris}
\affiliation{\it Department of Mathematics, Aberystwyth University, Ceredigion SY23 3BZ, Wales, UK}

\author{L. Truskinovsky}
\affiliation{\it  PMMH, CNRS -- UMR 7636, CNRS, ESPCI Paris, PSL Research University, 10 rue Vauquelin, 75005 Paris, France}

\date{\today}

\begin{abstract}
Energy dissipation by fast  crystalline defects  takes place  mainly through  the resonant  interaction  of their cores with   periodic lattice. We show that the resultant   effective  friction can be reduced to   zero  by appropriately tuned acoustic sources located on the boundary of the body.  To illustrate the general idea,  we  consider three  prototypical models describing the  main types of strongly discrete defects: dislocations, cracks and domain walls.  The obtained control protocols, ensuring  dissipation-free  mobility of topological defects, can be also used in the design of meta-material systems  aimed at transmitting mechanical information.
\end{abstract}
\maketitle

Mobile crystalline    defects respond  to lattice  periodicity   by  dynamically adjusting their core structure  which leads to   radiation of   lattice waves  through  parametric resonance~\cite{currie1977numerical,peyrard1984kink,kunz1985discrete,
boesch1989spontaneous,kevrekidis2002continuum}. Such 'hamiltonian damping' is one of the   main  mechanisms of energy loss for fast moving dislocations 
\cite{atkinson1965motion,kresse2004lattice}, crack tips \cite{slepyan1981dynamics,marder1995origin}  and  elastic phase/twin boundaries
\cite{slepyan2001feeding, truskinovsky2005kinetics}. 
Similar effective dissipation  hinders the mobility of topological \emph{defects} in mesoscopic dispersive systems, from  periodically
modulated composites\cite{dohnal2015dispersive}
to  discrete acoustic metamaterials \cite{kochmann2017exploiting}.

While at the macroscale  friction is usually diminished
  by applying lubricants,  at the microscale  it may be preferable to  use instead  external sources of 
ultrasound (sonolubricity) \cite{pfahl2018universal}. Correlated  mechanical vibrations are known to   reduce
\emph{macroscopic} friction through  acoustic ‘unjamming’ \cite{capozza2009suppression} as  in the case of 
 the remote triggering of earthquakes \cite{de2019induced}. General  detachment front tips serve as  macroscopic defects whose mobility in highly inhomogeneous environments can be controlled  by  AC (alternating current) driving  \cite{rubinstein2004detachment}.
Ultrasound-induced lubricity   can also reduce friction at the microscale 
 \cite{dinelli1997briggs}. It is known, for instance, that  the forming load 
  drops significantly in the presence of appropriately tuned time-periodic driving  which reduces  dislocation  friction~\cite{winsper1970introduction}.
 
The  AC-based  control  of the directed  transport  in \emph{damped} systems was studied extensively 
for the case when the sources  are  distributed in the bulk
 \cite{bonilla1991motion,cai1994moving,baizakov2007moving}. 
 In this Letter we  neglect  the conventional bulk dissipation, associated for instance,  with 'phonon
wind' \cite{koizumi2002lattice}, and show how in purely Hamiltonian setting  the effective  friction   can be  tuned  to zero by  the special AC driving  acting
   on the system  boundary \cite{tshiprut2005tuning,capozza2011stabilizing}. 
   
Since classical continuum models lack  the  resolution to describe
 dynamic  defect cores and therefore cannot capture  adequately the interaction between the defect and the external micro-structure,
we use 
atomistic models accounting for the 
coupling between the defect and the  lattice vibrations while respecting
the   anharmonicity of interatomic
forces. 
We build upon the theoretical methodology developed in \cite{slepyan2001feeding,mishuris2009localised, nieves2017transient} and show   that such   driving  can  compensate  radiative damping   \emph{completely},  making the discrete system   fully  transparent for  mobile topological  defects.  
   
   To highlight ideas we present a comparative study  of the three  prototypical snapping-bond type lattice models originating in   crystal plasticity (Frenkel-Kontorova (FK) model \cite{kresse2004lattice}), theory of structural phase transitions  (bi-stable Fermi-Pasta-Ulam (FPU) model \cite{efendiev2010thermalization})  and  fracture mechanics  (Peyrard-Bishop (PB) model, \cite{maddalena2009mechanics}).  
   
   In the individual setting of each of these models we study the effect  of the boundary AC sources  on  kinetic/mobility  laws for the corresponding lattice defects. The latter  relate  the macroscopic  driving force (dynamic generalization of the Peach-Koehler force in the case of dislocations,  the stress intensity factor in the case of cracks and the Eshelby force in the case of phase boundaries) and the velocity of the defect. We find  that in the presence of AC sources such relations  becomes \emph{multivalued}. We focus particularly on designing the  AC protocols which ensure that  the  steady propagation of a defect   takes place  under \emph{zero} driving force. 
     
    The possibility of  externally guided    radiation-free propagation of  mechanical information is presently  of considerable interest for designing   discrete meta-materials  with buckling linkages.  Geometric phase transitions generating information-carrying defects  in such systems  play a central role in a multitude of new applications  from  recoverable energy harvesting to  controlled structural collapse \cite{shan2015multistable,zhang2019programmable,harne2017harnessing}.


\begin{figure}[!htbp]
\includegraphics[scale=.7]{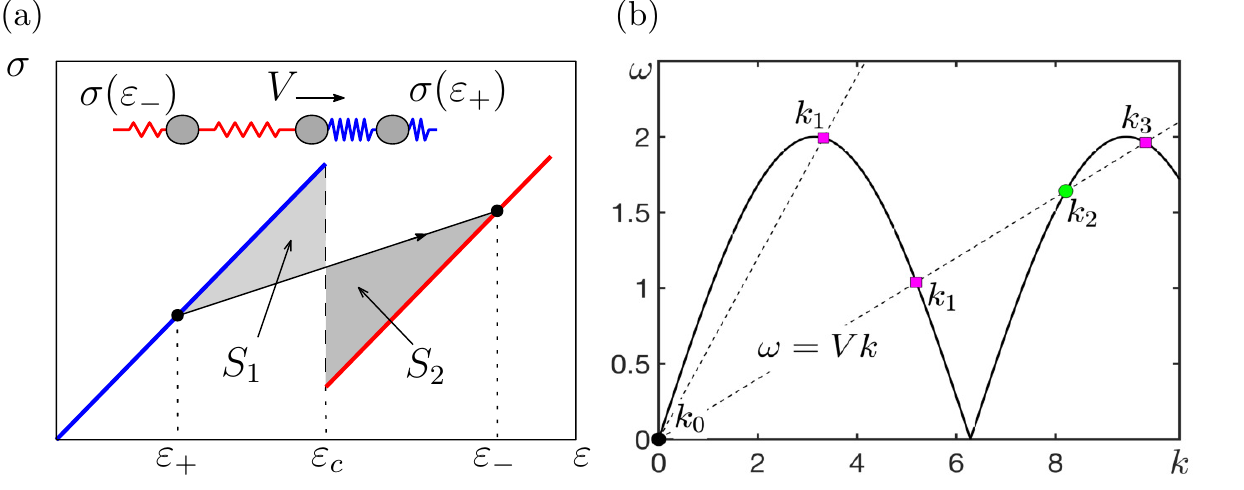}
\caption{(a)  Piece-wise linear stress-strain relation $\sigma=\sigma(\varepsilon)$;   the macroscopic driving force $G^M(V)=S_2-S_1$. (b) Dispersion relation $\omega(k)$  for $\text{Im}(k)=0$ (acoustic branch);  $k_j$ correspond to  the radiated   waves   in cases   $K=1$ and $ K=3$, see the text. Green circles correspond to  AC sources behind the defect,   magenta squares - ahead  of the  defect.}  
\label{fig:chain}
\end{figure}

 The FPU model with bi-stable interactions is used  to represent  the simplest crystal defect,  a domain wall \cite{truskinovsky2005kinetics, slepyan2005transition}.
In   terms of dimensionless particle displacements $u_j(t)$ the dynamics is described by the   system   \begin{equation}
 \ddot{u}_j(t) =\sigma\left( u_{j+1}-u_{j} \right)-\sigma\left( u_{j}-u_{j-1} \right). \label{dyn}
 \end{equation}
It will be  convenient to  use strain variables  $\varepsilon_j(t)=u_{j+1}(t)-u_j(t)$ and  introduce  the strain energy density $w(\varepsilon)$ so that  $\sigma(\varepsilon_j)=w'(\varepsilon_j)$. For analytical transparency we adopt   the simplest bi-quadratic model with
 $w=
 (1/2)\varepsilon^2-\sigma_0(\varepsilon-\varepsilon_c) H(\varepsilon-\varepsilon_c)$,  where $H(x)$ is the Heaviside function, 
  $\varepsilon_c$ is the  characteristic  strain and $\sigma_0$ is the   stress drop, see  Fig.~\ref{fig:chain}(a).

We search for traveling wave  (TW)  solutions of \eqref{dyn} in the form $ u_j(t)= u(\eta)$  and  $\varepsilon_j=\varepsilon(\eta)$,  where $\eta=j-Vt$  and  $V<1$ is the  normalized velocity of the defect.  If we associate  the defect  with    $\eta=0$  the   equation for the strain field  reduces to
$
V^2 d^2\varepsilon/d \eta^2=\sigma(\eta+1)+\sigma(\eta-1)-2\sigma(\eta),
$
where  $
\sigma(\eta)=\varepsilon(\eta)-\sigma_0 H(-\eta).
$
When this linear equation is solved, the  velocity $V$ is found from the nonlinear \emph{switching} condition  $\varepsilon(0)=\varepsilon_c$.
 
 Using the Fourier transform $\hat{f}(k)=\int_{-\infty}^{\infty}f(\eta)e^{ik\eta}\,d\eta$,  we can rewrite the main linear problem in the form  $L(k) \hat{\varepsilon}(k)=\sigma_0 \omega^2(k)/(0+ik)$, where $L(k) \equiv \omega^2(k)-(kV)^2 $ and $\omega^2(k)=4\sin^2{(k/2)}$ is the 
dispersion relation  represented in this case by a single  acoustic branch, see  Fig.~\ref{fig:chain}(b). 
The strain field $\varepsilon (\eta)$ can be decomposed into a sum of the term $\varepsilon_{in}(\eta)$, which is due to inhomogeneity (mimicking nonlinearity) 
and the  term  $\varepsilon_{dr}(\eta)$, due to the combined action of DC (direct current) and AC  driving.  
The former  can be written explicitly
\begin{equation}
 \varepsilon_{in}(\eta)=\frac{\sigma_0}{2\pi}\int_{-\infty}^{\infty}\frac{\omega^2(k)e^{-ik\eta}}{(0+ik)L(k)}\,dk.
\label{eq:Solution0}
\end{equation}
The latter must satisfy $L(k)\hat{\varepsilon}_{dr}(k)=0$ which   in the physical space  gives 
\begin{equation}
\varepsilon_{dr}(\eta)=
 \sum_{j=1}^K A_j\sin{(k_j\eta + \varphi_j)} + C . 
\label{eq:Solution}
\end{equation}
The constants  $A_j$ and $\varphi_j$  describe the amplitude and the phase of the incoming waves generated at  the  distant boundaries.  They  represent the AC driving which is characterized by the wave numbers  $k_j$ that are taken among  the positive real roots of the kernel  $L(k)$: if $\omega'(k_j)$ is smaller (greater) than $V$ the sources are in front of (behind) the moving defect.  The constant $C$ in \eqref{eq:Solution}, representing the   root  $k_0=0$,   controls  the uniform  strain ahead of the moving defect and represents  the  DC driving.  

Use the switching  condition  
we   can obtain  the explicit relations for the limiting strains  in the form 
 $\langle \varepsilon \rangle(\pm \infty)\equiv\varepsilon_\pm=\varepsilon_c\mp\frac{1}{2}\frac{\sigma_0}{1-V^2}+\sigma_0 Q-\sum_{j=1}^KA_j\sin{\varphi_j},$ where  
 $\langle f \rangle=\lim_{T\to\infty}(1/T)\int_0^{T}f(s)ds$ and  the   expression for the universal function $
Q(V)$ can be found  in \cite{SOM}.
It can be checked that the obtained solution respects  the macroscopic 
 momentum balance represented by one of the  
 Rankine-Hugoniot (RH) conditions \cite{dafermos2005hyperbolic}: $
V^2=(\sigma(\varepsilon_+)-\sigma(\varepsilon_-))/(\varepsilon_+-\varepsilon_-)$. 
The limiting values of   the mass velocity $v_j=\dot{u}_j$
naturally  satisfy another (kinematic)  RH condition 
$\langle v \rangle(\pm \infty) \equiv v_\pm=-V  \varepsilon_{\pm}$. 
 


We now write  the macroscopic energy dissipation on the moving defects as  $\mathcal{R}=VG\geq 0$ where $G$ is the driving force. In the absence of  the AC driving ($A_j=0$) we obtain $G=G^M $, where 
\begin{equation}
G^M=\llbracket {w} \rrbracket-\{\sigma\}\llbracket \varepsilon \rrbracket,
\end{equation}
and we used the standard notations   $\llbracket f \rrbracket =f_+-f_-$ and $\{ f\}=(f_++f_-)/2$ \cite{truskinovskii1987dynamics}. In our case   $
G^M=(\sigma_0/2)(\varepsilon_++\varepsilon_--2\varepsilon_c)$. With the AC driving present,  we need to write  $\mathcal{R}=V(G^M+G^{m}) \geq 0$   where
 the total power exerted by microscopic sources  is
 \begin{equation}
VG^{m}=\sum_{j=1}^K (1/2)A_j^2\left| \omega'(k_j)-V\right|\geq 0.
\end{equation}
The relation for $\mathcal{R}$ can be checked by the independent computation of the energy carried   by the microscopic radiation away from the moving defect to infinity \cite{SOM}.
 \begin{figure}[!htbp]
\includegraphics[scale=0.67]{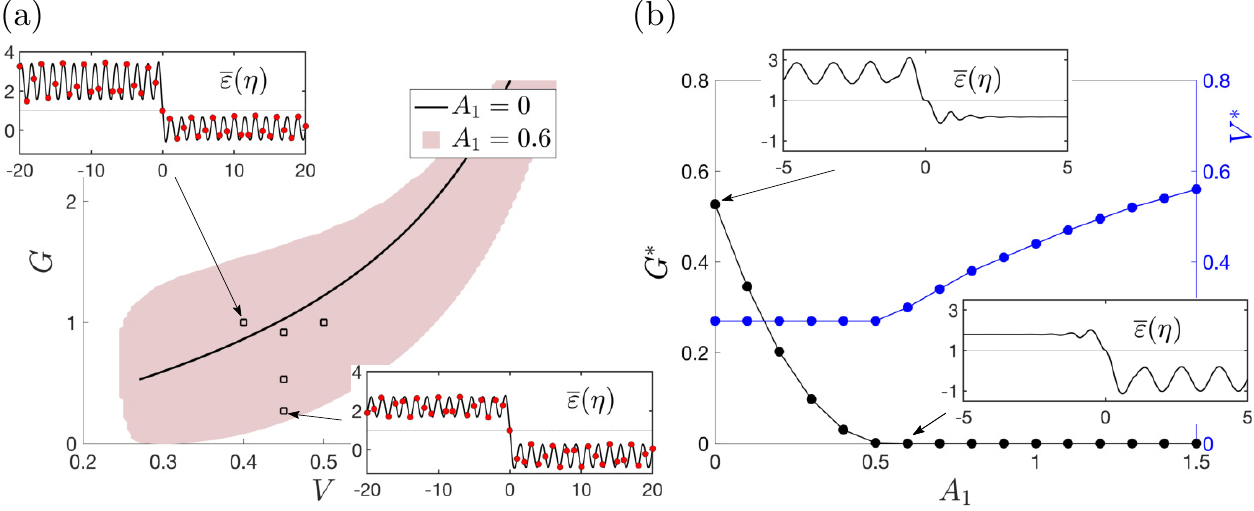}
\caption{(a) Kinetic domain for the case  $K=1$, open squares show the selected TW solutions reached numerically  \cite{SOM}; 
(b) Amplitude dependence of $G^*$ and  $V^*$; insets show normalized  strains $\overline{\varepsilon}(\eta)=\varepsilon(\eta)/\varepsilon_c$. Parameters: $\sigma_0=2,\,\varepsilon_c=1$.}
\label{fig:G_min_V}
\end{figure}

 The  dependence  of  $G$ on $V$ for a high velocity subset of admissible solutions  is  shown in Fig.~\ref{fig:G_min_V}(a).  The  radiative damping is represented here by a single  wave  number $k_1$. The AC driving is tuned to the same wave number and its source is placed ahead of the moving defect (the $K=1$ regime).  If the AC driving is absent  and all $ A_j=0$, there is a single value of $V$ for each value of  $G$ within the admissible range  $ \pm(\varepsilon_c-\varepsilon(\eta))>0$ at $\pm\eta>0$. Even if only one coefficient  $A_1 \neq 0$,  each admissible value of velocity $V$  can be reached within  a finite range  of DC driving amplitudes with  the associated phase shift $\varphi_1$ varying continuously.  In this case   the   \emph{kinetic relation }  transforms   into a 2D  \emph{kinetic domain}, see Fig.~\ref{fig:G_min_V}(a), where  by fixing the DC drive  we can either speed up or slow down the defect   as we  change the frequency of the AC source.

\begin{figure}[!htbp]
\includegraphics[scale=0.7]{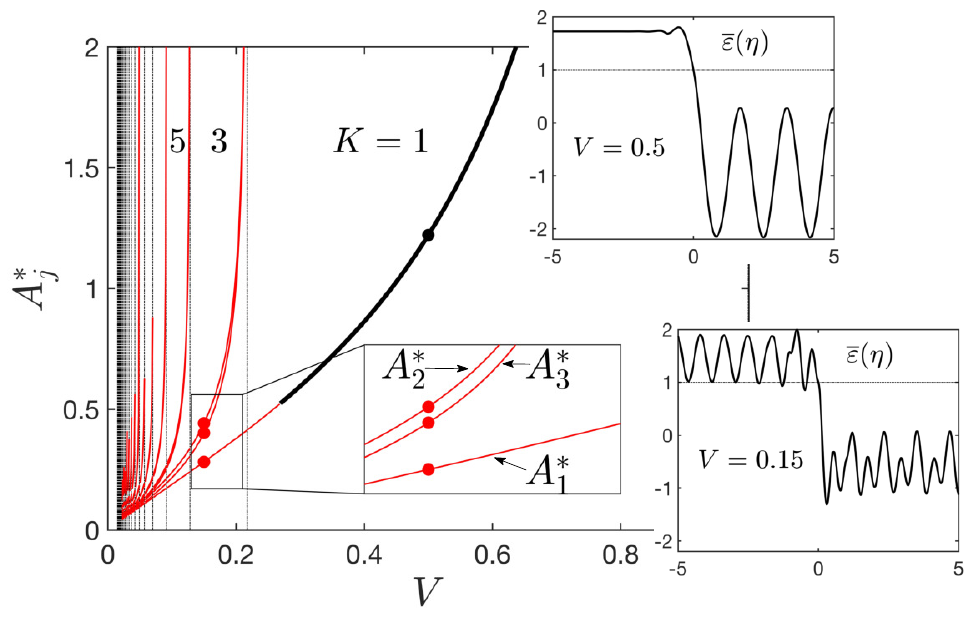}
\caption{Amplitudes of the AC  sources $A_j^*(V),\, j=1,2,..,K$ for $K=1,3,5...$; insets show strains $\overline{\varepsilon}(\eta)=\varepsilon(\eta)/\varepsilon_c$ at $V=0.15$ and $V=0.5$ with corresponding $A_j^*$ marked by solid circles. Black lines correspond to admissible solutions, red - to non admissible. Parameters: $\sigma_0=2,\,\varepsilon_c=1$.}
\label{fig:A_star}
\end{figure}

The  possibility of the AC induced   friction  \emph{reduction} is seen  from the fact that for each  $V$ there is  a range of the admissible   driving forces  $G$ with the   minimal value  $G^*(V)$.
Moreover,   for  some $V^*$ such  friction can be  eliminated completely if the amplitude $A_1$ reaches beyond a  threshold. Note that the emergence of friction-free  regimes resembles a second order phase transition with the dissipation $G^*$ as the order parameter, see Fig.~\ref{fig:G_min_V}(b). The  non-dissipative regimes with $K=1$ are   naturally  anti-phase with respect to the radiated waves  so that  $\varphi_1^*=\pi/2$, see a typical strain distribution  in the insets in   Fig.~\ref{fig:G_min_V}(b) and Fig. \ref{fig:A_star}. The   relation between  the AC amplitude and  the defect velocity  for such regimes can be written explicitly
$
A_1^*= \sigma_0V/(V-\omega'(k_f)).
$

In the general case $K\neq1$  the number of dissipative  waves is  odd and the  dissipation-free  regimes    also must  have an odd number of AC sources to cancel each of these waves.  Consider, for instance, the case   $K=3$,  illustrated in Fig. \ref{fig:chain}(b),  where  two dissipative lattice waves ($k_1$ and $k_3$)  release energy at  $-\infty$ and one wave ($k_2$) - at  $+\infty$. 
To block  these  dissipative waves one must have the sources of  AC driving both in front and behind the  defect. The corresponding   amplitudes, ensuring that 
  $G^*(V)=0$,  are $A^*_{j}=(-1)^j \sigma_0V/(\omega'(k_j)-V)$ with $j=1,2,3$, see Fig.~\ref{fig:A_star}.

The numerical check  shows that the  admissible frictionless regimes
exist only for $K=1$. To show numerical stability of these regimes  we   simulated the transient problem with   initial data close to the analytical TW solutions, see  \cite{SOM} for details. The simulation involving $1000$ equations and showing  stable dissipation-free propagation of the defect is presented in the form of  the supplementary Movie 2.

\begin{figure}[!htbp]
\includegraphics[scale=.7]{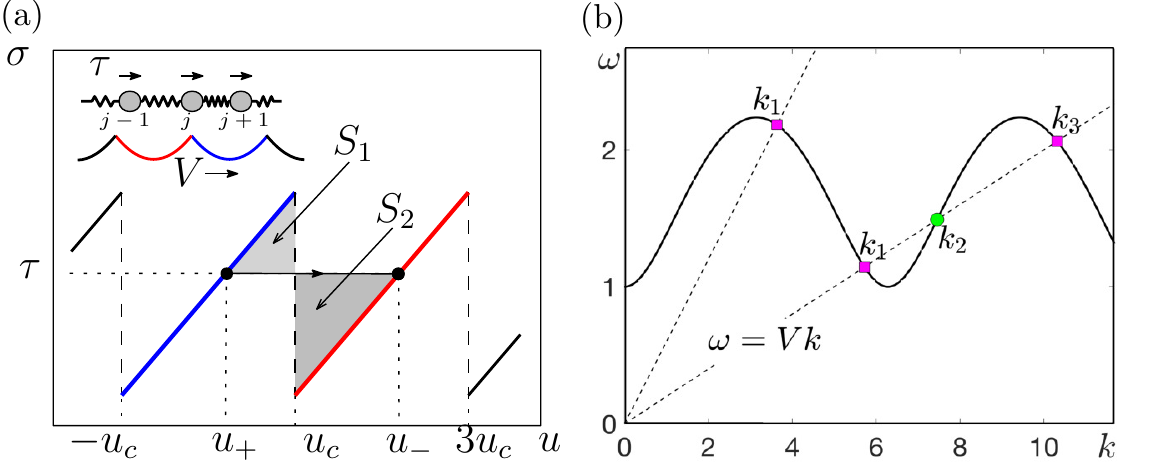}
\caption{(a) Dislocation propagation driven by the  constant force $\tau$; macroscopic driving force $G^M(V)=S_2-S_1$;  (b) Dispersion relation for $\text{Im}(k)=0$ (optical  branch). The wave numbers $k_j$ define radiated   waves in the cases $K=1$ and $ K=3$.}
\label{fig:chain_dislocation}
\end{figure}

 The simplest FK model \cite{atkinson1965motion,kresse2003mobility}  can be used to analyze the frictionless propagation regimes for moving dislocations, see  Fig.~\ref{fig:chain_dislocation}(a).   To describe a single dislocation  we only need two wells of the on-site periodic potential.  The displacement $u_{j}(t)$,    describing  horizontal  slip,  must solve the equations \begin{equation}
\ddot{u}_j=u_{j-1}+u_{j+1}-2u_{j}-\sigma(u_j)+\tau
\end{equation}
 where $\tau$ is a uniform load. The function $\sigma(u)$ is illustrated in Fig. \ref{fig:chain_dislocation}(a)  and is defined via the on-site potential $w(u)=(1/2)u^2$ when $-u_c<u<u_c$ and $w(u)=(1/2)u^2-\sigma_0(u-u_c)$ when $u_c<u<3u_c$  representing the two relevant periods.
  We   again use  the TW ansatz $u_j(t)=u(\eta),\, \eta=j-Vt$ and apply the corresponding condition of admissibility.
Unlike the previous case, the DC drive $\tau$ is now applied in the bulk. 
\begin{figure}[!htbp]
\includegraphics[scale=0.7]{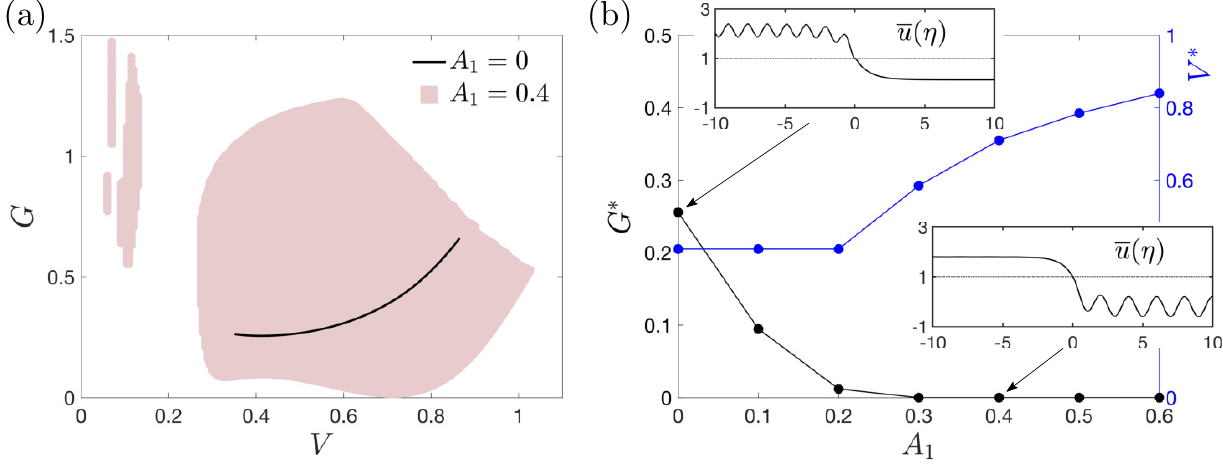}
\caption{(a) Kinetic domains for dislocations showing admissible solutions with $K=1$. There is only one damping wave in the big pink domain and more than one  in the smaller ones. (b) AC amplitude dependence of $G^*(V^*)$ and $V^*$; insets show displacements $\overline{u}(\eta)=u(\eta)/u_c$. Parameters: $\sigma_0=2,\,u_c=1$.}
\label{fig:G_dislocation}
\end{figure}

We need to solve the linear equation $V^2 u''(\eta)=u(\eta+1)+u(\eta-1)-3u(\eta)+\sigma_0H(-\eta)+\tau$ and then find the defect velocity $V$ using the nonlinear \emph{switching} condition $u(0)=u_c$. The solution can be again represented in the form $u(\eta)=u_{in}(\eta)+u_{dr}(\eta)$. The first term, which is due to inhomogeneity (mimicking nonlinearity)  now  includes the DC driving  $\tau$: 
\begin{equation}
u_{in}(\eta)=\tau+\frac{\sigma_0}{2\pi}\int_{-\infty}^{\infty}\frac{e^{-ik\eta}}{(0+ik)L(k)}\,dk,
\end{equation}
where the operator $L(k)$ remains  the same as in the FPU problem but the dispersion relation  $\omega^2(k)=4\sin^2(k/2)+1$ is now represented by a single  optical branch.
The second term responsible for the AC driving  must again satisfy   $L(k)\hat{u}_{dr}(k)=0$ and can be again represented    as a combination of linear  waves whose phase velocity is equal to $V$
\begin{equation}
u_{dr}(\eta)=\sum_{j=1}^K A_j\sin(k_j\eta+\varphi_j).
\end{equation} 
Here  $k_j$ are again the positive real roots of $L(k)=0$.

Using  the  switching  condition  we obtain for the time averaged displacements at $\pm \infty$ the  values 
$u_\pm=u_c\mp(\sigma_0/2)+\sigma_0R-\sum_jA_j\sin\varphi_j$ 
with   
the explicit expression for the universal function $R(V)$  is given again in \cite{SOM}. 
The analogs of the RH conditions are now  $u_+=\tau$ and  $u_-=\tau+\sigma_0$. 

If we denote the stress in the horizontal bonds by $\bar{\sigma}(\varepsilon)=\varepsilon$ we can write  the  rate of  dissipation at  the macro-scale as $VG^M=\llbracket v^2/2+\varepsilon^2/2+w-\tau u \rrbracket V+\llbracket \bar{\sigma} v\rrbracket $. Applying the kinematic RH condition $\llbracket  v \rrbracket+V\llbracket \varepsilon \rrbracket=0$ 
 we obtain 
 \begin{equation}
 G^M=\llbracket w \rrbracket -\tau \llbracket u\rrbracket.
 \end{equation}
Since  $\varepsilon_\pm=\bar{\sigma}(\varepsilon_\pm)=0$ we can finally write the macroscopic driving force in the form  $
 G^{M}=\sigma_0^2/2-\sigma_0(u_c-u_+).
$
  The contribution to the energy flux due to  AC sources is now
  \begin{equation}
  VG^{m}=\sum_{j=1}^K (1/2)A_j^2\omega^2(k_j)\left| \omega'(k_j)-V\right| \geq 0.
 \end{equation}
 The  multi-valued relation  $\mathcal{R}(V) =VG(V)= G^{M}(V)+G^{m}(V)$ for admissible solutions is illustrated   in  Fig.~\ref{fig:G_dislocation}(a) for the case $K=1$. 
It is again possible to completely cancel the  lattice friction  and obtain regimes with   $G^*(V^*)=0$.   In such regimes, illustrated  for $K=1$ in Fig.~\ref{fig:G_dislocation}(b),  the  radiated waves are again  annihilated  by  the waves generated by  the  AC source with the amplitudes $A_j^*=(-1)^j\sigma_0V/(\omega(k_j)^2(\omega'(k_j)-V))$ and phase shifts $\varphi^*_j=\pi/2$. 
 
\begin{figure}[!htbp]
\includegraphics[scale=.7]{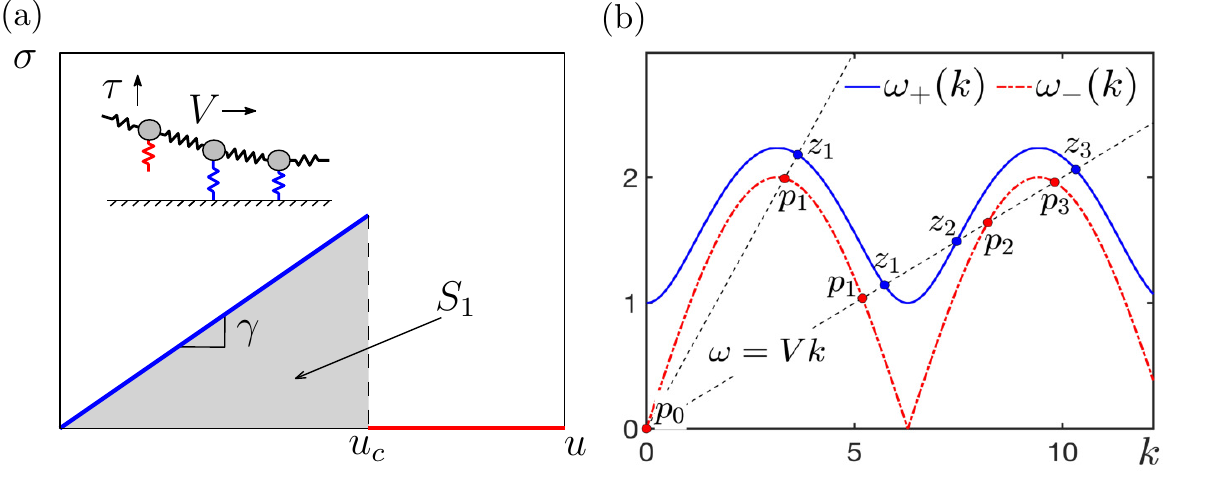}
\caption{(a) Constitutive relation for a mechanical fuse and schematic representation of a crack propagating with velocity  $V$ under remote load $\tau$; the strain energy jump $\llbracket w\rrbracket=-S_1$. (b) Dispersion relations $\omega_+(k)$ (optical branch) and $\omega_-(k)$ (acoustic branch)  for $\text{Im}(k)=0$ characterizing  intact and broken lattices, respectively.}
\label{fig:chain_fracture}
\end{figure}

Our last example deals with reversible fracture in the simplest PB-type setting  \cite{peyrard1989statistical,marder1995origin, maddalena2009mechanics}. The lattice defect is now    a  crack tip moving under the action of a transversal force from left to right by  consequently breaking the bonds represented by elastic fuses, see  Fig.~\ref{fig:chain_fracture}(a).  

 The equations governing the evolution of the  vertical displacements $u_j(t)$ are
\begin{equation}
\ddot{u}_j=u_{j-1}+u_{j+1}-2u_{j}-\gamma u_jH(u_j-u_c)
\end{equation}
see  Fig.~\ref{fig:chain_fracture}(a) for notations,  and we again look for  solutions  in the TW form $u_j(t)=u(\eta),\,\eta=j-Vt$. We need to solve a linear equation  $V^2u''(\eta)=u(\eta+1)+u(\eta-1)-2u(\eta)-\gamma u(\eta)H(\eta)$ and use  the nonlinear  \emph{switching} condition $u(0)=u_c$ to find the defect velocity $V$. The dispersion relations are   now represented by one optical branch $\omega_{+}^2(k)=4\sin^2{(k/2)}+\gamma$ ahead and one acoustic branch $\omega_{-}^2(k)=4\sin^2{(k/2)}$ behind the defect.


One way to solve  this more complex problem  is to use the Wiener-Hopf technique, see \cite{SOM} for details. We can again obtain  the decomposition $u(\eta)=u_{in}(\eta)+u_{dr}(\eta)$, but now to define  different terms we need to introduce two auxiliary functions
$ 
L^\pm(k)=L^{\mp 1/2}(k)\exp{(\frac{1}{2\pi i} \int_{-\infty}^{\infty}\frac{\text{Log }L(\xi)}{k-\xi}d\xi)},
$  where
 $L(k)\equiv (\omega_+^2(k)-(kV)^2)/(\omega_-^2(k)-(kV)^2)$. 
Then
\begin{equation}
u_{in}(\eta)=\frac{C}{2\pi}\int_{-\infty}^{\infty}\frac{L^\pm(k)e^{-ik\eta}}{0\mp ik}\,dk,\quad \pm \eta>0
\label{1}
\end{equation}
is the contribution due to  remotely applied DC force $\tau$ which is  modeled by the condition that at $\eta=-\infty$ the time average displacements follows the asymptotics $ u(\eta) \sim -\tau \eta$, while at $\eta=+\infty$ the average displacements tend  to zero. From these conditions  we find that  $C=\tau S \sqrt{(1-V^2)/\gamma}$ where  an explicit expression for the function $S(V)$ is given in \cite{SOM}. The contribution due to the  AC driving is
\begin{equation}
u_{dr}(\eta)=\frac{1}{2\pi}\int_{-\infty}^{\infty}L^\pm(k)\Psi_{dr}^\pm(k)e^{-ik\eta}\,dk,\quad \pm\eta>0
\label{1}
\end{equation}
where
\begin{equation}
\Psi_{dr}^\pm(k)=
\sum_{j=1}^K \frac{A_j}{2}\left[\frac{e^{-i(\varphi_j-\pi/2)}}{0\mp i(k-k_j)}+\frac{e^{i(\varphi_j-\pi/2)}}{0\mp i(k+k_j)}\right].
\label{psi}
\end{equation}
Here   the wave numbers $k_j=z_{2j-1}$   describe  the sources bringing the energy   from $+\infty$ while the wave numbers  $k_j=p_{2j}$ correspond to sources bringing the energy from $-\infty$; for the cases $K=1,3$ the wave numbers $z$ and $p$ are illustrated in Fig.~\ref{fig:chain_fracture}(b).

\begin{figure}[!htbp]
\includegraphics[scale=0.7]{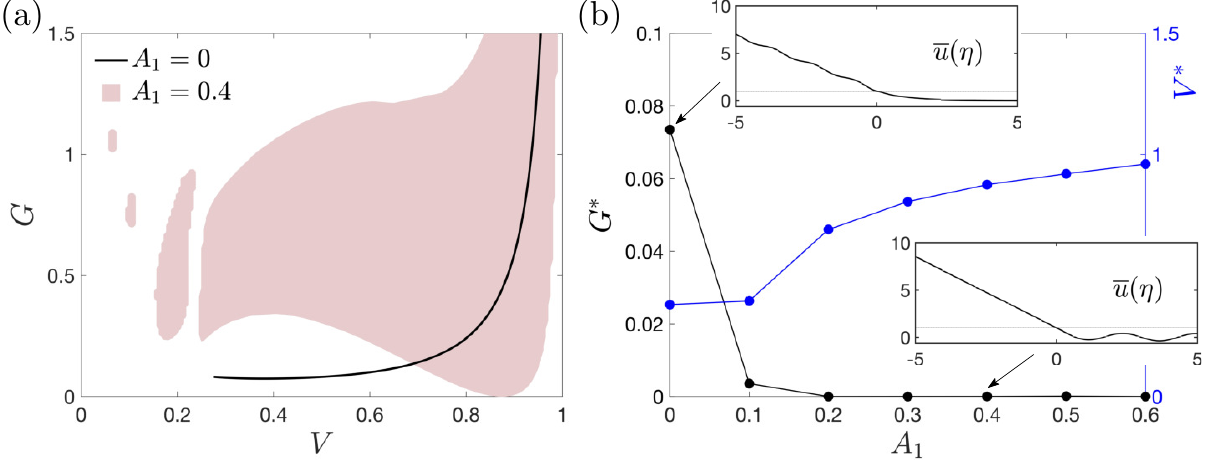}
\caption{(a) Kinetic domains for admissible solutions with $K=1$ and the AC wave coming from ahead. There is only one damping wave in the big pink domain and more than one  in the smaller ones.  (b) Amplitude dependence of $G^*(V^*)$ and $V^*$. Material properties are $\gamma=1,\,u_c=1$.}
\label{fig:G_crack}
\end{figure}

To compute the driving force $G^M$ we observe that the  macroscopic  energy dissipation on the crack tip  is
$
VG^M=\llbracket  v^2/2+\varepsilon^2/2+w \rrbracket V+
\llbracket \bar{\sigma} v \rrbracket$, where $w(u)=(1/2)\gamma u^2$ for  $u<u_c$ and $w(u)=(1/2)\gamma u_c^2$ for $u>u_c$.  
If we now take into consideration the RH compatibility condition
$\llbracket  v \rrbracket+V\llbracket \varepsilon \rrbracket=0$, we obtain 
$VG^M=\llbracket w \rrbracket V+\bar f \{v\}$. Here $\bar f=\llbracket  \bar{\sigma} \rrbracket -V^2
\llbracket \varepsilon \rrbracket$ is the moving concentrated force which represents  the microscopic  processes in the tip and furnishes  the linear momentum RH condition, e.g.~\cite{burridge1978peeling}. We can now write
\begin{equation}
G^M=\llbracket w \rrbracket  -\bar f \llbracket \varepsilon \rrbracket
\end{equation}
 and 
substituting the values $\varepsilon_-=-\tau$, $\varepsilon_+=0$,
$
\bar{\sigma}(\varepsilon_{\pm})=\varepsilon_{\pm},
$
we finally obtain
$
G^M=\tau^2(1-V^2)/2-(\gamma u_c^2)/2.
$

Consider the simplest case when there is only one radiated wave with $k_1=p_1$. The microscopic power exerted by a single  AC source   ahead of the crack ($K=1$)
is then \begin{equation}
VG^{m}=\frac{1}{2} A_1^2\omega^2_+(z_1)|L^+(z_1)|^2\left | \omega_+'(z_1) -V\right |  \geq 0. 
\end{equation}
The total dissipation  $\mathcal{R}(V) =VG(V)=V(G^{M}(V)+G^m(V)) \geq 0$ is again a multivalued function of $V$  as we show  in Fig.~\ref{fig:G_crack}(a);  the associated  functions $G^*(V^*)$ and $V^*$ at different values of $A_1$ are shown in Fig.~\ref{fig:G_crack}(b). 
  At a given $V$ we obtain  $A^*_1=u_c(z_1^2-p_1^2)/z_1^2$ and  $\varphi^*_1=\pi/2$ with the corresponding dissipation-free solution illustrated in the inset in Fig.~\ref{fig:G_crack}(b).
More general solutions,  similar to the ones in Fig. \ref{fig:A_star}, can be  obtained as well. 

To conclude, we  showed that it is possible to  fine tune  defect kinetics by carefully engineered AC driving. Moreover, using special AC sources on the boundary,  one can  compensate radiative damping completely,   
making the crystal free of internal friction  for strongly discrete defects. We demonstrated this effect  for  domain boundaries, dislocations and cracks, however,    the obtained  results  also have important implications  for the design of  artificial metamaterials  supporting  mobile  topological defects and capable of transporting  compact units of mechanical  information.  

 

\end{document}



\title{Supplementary material for the paper: "Frictionless motion of lattice defects"}

\author{N.Gorbushin}
\affiliation{\it  PMMH, CNRS -- UMR 7636, CNRS, ESPCI Paris, PSL Research University, 10 rue Vauquelin, 75005 Paris, France}

\author{G. Mishuris}
\affiliation{\it Department of Mathematics, Aberystwyth University, Ceredigion SY23 3BZ, Wales, UK}

\author{L. Truskinovsky}
\affiliation{\it  PMMH, CNRS -- UMR 7636, CNRS, ESPCI Paris, PSL Research University, 10 rue Vauquelin, 75005 Paris, France}

\date{\today}

\maketitle
\section{FPU problem: analytical results}

In this problem the  characteristic function $L(k)$ has the following properties: $L(-k)=L(k)$ and $L(\overline{k})=\overline{L(k)}$ and, hence, the  real and purely imaginary roots of the characteristic equation $L(k)=0$ come in pairs while  other complex roots come in quadruplets. Therefore, it is enough to search for the roots of  the characteristic equation  in the quarter of the complex plane $\text{Im }k\geq 0$ and $\text{Re }k\geq 0$. The associated roots can be conveniently sorted between the sets $Z^\pm=Z_c^\mp \cup Z_r^\pm$ with 
$Z^\pm_c=\left\{k:\,L(k)=0,\, \pm \text{Im }k>0\right\}$ and 
$Z^\pm_r=\left\{|k|>0\,:\,L(k)=0,\,\text{Im }k=0,\, \pm kL'(k)>0\right\}
$. 

With the roots known, the integration in Eq. 2 in the main text can be performed explicitly which allows one to  obtain   the expression for the full strain field:
\begin{equation}
\varepsilon(\eta)=\begin{cases}
  \varepsilon_+ + \sum_{j=1}^K A_j\sin{(k_j\eta + \varphi_j)} -  \sum\limits_{k_j \in Z^+}\dfrac{\sigma_0\omega^2(k_j)}{k_j L'(k_j)}e^{-ik_j\eta},\, \eta>0,\\
 \varepsilon_- + \sum_{j=1}^K A_j\sin{(k_j\eta + \varphi_j)} +\sum\limits_{k_j \in Z^-}\dfrac{\sigma_0\omega^2(k_j)}{k_j L'(k_j)}e^{-ik_j\eta},\, \eta<0,
\end{cases}
\label{eq:Solution}
\end{equation}
To  use the switching condition  $\varepsilon(0)=\varepsilon_c$, we  need to recall  the following   general properties of the roots of the characteristic equation~\cite{truskinovsky2005kinetics}
\begin{equation}
\sum\limits_{k_j \in Z_c^+}\dfrac{\omega^2(k_j)}{k_j L'(k_j)}+\sum\limits_{k_j \in Z_c^-}\dfrac{\omega^2(k_j)}{k_j L'(k_j)}=-\frac{1}{1-V^2}-\sum\limits_{k_j \in Z_r^+}\dfrac{\omega^2(k_j)}{k_j L'(k_j)}-\sum\limits_{k_j \in Z_r^-}\dfrac{\omega^2(k_j)}{k_j L'(k_j)},
\end{equation}
\begin{equation}
\sum\limits_{k_j \in Z_c^+}\frac{\omega^2(k_j)}{k_j L'(k_j)}=\sum\limits_{k_j \in Z_c^-}\frac{\omega^2(k_j)}{k_j L'(k_j)}.
\end{equation}
We can rewrite the switching condition in two equivalent forms
%
$
\varepsilon_{\pm}=\varepsilon_c\mp\frac{\sigma_0/2}{1-V^2}+\sigma_0Q+ \sum_{j=1}^K A_j\sin{\varphi_j},
$
where 
\begin{equation}
Q=\frac{1}{2}\sum\limits_{k_j \in Z_r^+}\frac{\omega^2(k_j)}{k_j L'(k_j)}-\frac{1}{2}\sum\limits_{k_j \in Z_r^-}\frac{\omega^2(k_j)}{k_j L'(k_j)}.
\end{equation}

Next,  we compute  the rate of dissipation   by lattice waves:
\begin{equation}
{\cal R}_\pm=\sum_{k_j\in Z_r^\mp}\langle {\cal E}_j \rangle |\omega'(k_j)-V|,
\end{equation}
where ${\cal E}_j=v_j^2/2+w(\varepsilon_j)$ is the   energy density carried by the 
linear wave with the (real) wave number $k_j>0$ and $\omega'(k_j)-V$ is the velocity of the  energy drift relative  to the velocity of the defect; the signs indicate waves carrying the energy to $\pm \infty$.  The particle velocity here $v(\eta)=-Vdu/d\eta$ can be obtained  by inverting the kinematic relation in the Fourier space $\hat{v}(k)=-Vk\exp{(ik/2)}/[2\sin{(k/2)}]\hat{\varepsilon}(k)$. We obtain explicitly:
\begin{equation}
v(\eta)=\begin{cases}
  -V\varepsilon_+ - \sum_{j=1}^K \dfrac{A_jVk_j}{2\sin{(k_j/2)}}\sin{(k_j(\eta-1/2) + \varphi_j)} +  \sum\limits_{k_j \in Z^+}\dfrac{\sigma_0Vk_j\omega^2(k_j)}{2k_j\sin(k_j/2)L'(k_j)}e^{-ik_j(\eta-1/2)},\, \eta>1/2,\\
 -V\varepsilon_- - \sum_{j=1}^K \dfrac{A_jVk_j}{2\sin{(k_j/2)}}\sin{(k_j(\eta-1/2)+ \varphi_j)} -\sum\limits_{k_j \in Z^-}\dfrac{\sigma_0Vk_j\omega^2(k_j)}{2k_j\sin(k_j/2)L'(k_j)}e^{-ik_j(\eta-1/2)},\, \eta<1/2.
\end{cases}
\label{eq:Solution_v}
\end{equation}
After the substitution we obtain 
\begin{equation}
G_\pm=\frac{{\cal R}_\pm}{V}=\sum_{k_j\in Z_r^\pm,k_j>0}\left[\left(\pm2\frac{\sigma_0\omega^2(k_j)}{k_jL'(k_j)}-A_j\sin\varphi_j\right)^2+A_j^2\cos^2{\varphi_j}\right]\left|\frac{\omega'(k_j)}{V}-1\right|.
\label{eq:EnergyFluxes}
\end{equation}
The first term in the square parenthesis reveals the interaction between the waves generated by AC forces and the  waves radiated  by the defect. The second term corresponds to the contribution from the AC sources  only. If we use the definitions of $G^M$ , Eq. 4, and $G^m$, Eq. 5,  in the main text and substitute 
the expressions of the fields $\varepsilon(\eta)$ and $v(\eta)$, we obtain the relation
%
\begin{equation}
G^M+G^m=G_+ + G_-.
\label{eq:EnergyBalance}
\end{equation}
To verify this identity it is enough to observe that $2\omega^2(k_j)/(k_jL'(k_j))=V/(\omega'(k_j)-V)$. 
From the representation \eqref{eq:EnergyFluxes} it is particularly easy to conclude that the condition $G^M+G^m=0$ is satisfied if we set $\varphi_j=\varphi_j^*=\pi/2$ and amplitudes $A_j=A_j^*=(-1)^j\sigma_0V/(\omega'(k_j)-V),\,j=1,2,...,K$.

\section{FPU problem: numerical  experiments}

To show how the DC and AC driving can be actually implemented and to show stability of the obtained analytical solutions we
we conducted  a series of direct numerical experiments  with a  finite chain comprised of $N=1001$ masses connected by bi-stable springs.

Our initial conditions contained  a pre-existing defect (phase boundary) located  at $n_0=200$. We assigned initial  displacements  (linear with $j$) in the form $u_j(0)=\tilde{\varepsilon}_-j$ for $j\leq n_0$ and $u_j(0)=\tilde{\varepsilon}_-n_0+\tilde{\varepsilon}_+(j-n_0)$ for $j>n_0$,  where $\tilde{\varepsilon}_\pm=\varepsilon_c\mp \sigma_0/2+\varepsilon_+-A/2$. Here and below, constants $\varepsilon_\pm$  correspond to the  limiting strains and $A$ is an amplitude of the AC drive which we use in the analytical solution. The initial velocities are set to 0 for each mass: $\dot{u}_j(0)=0$.



We fix the end of the chain on the right  side by setting: $u_{N+1}(t)=u_N(0)+\tilde{\varepsilon}_+$.  The left end is loaded by a constant force $F=\sigma(\varepsilon_-)+\sigma_0V/(2(1+V))$  representing the DC driving so that:  $\ddot{u}_1(t)=\sigma(u_1-u_2)-FH(t-t_0)$.

The time shift   $t_0$ is  needed for the energy from the AC source to arrive to the defect located at   $n=n_0$. 
We associate the AC driving source  with the two neighboring masses  $j_0=400$ and $j_0=401$ which are located sufficiently far ahead of the initial defect  and the right end of a chain. More specifically, we assume that a time periodic  pair of force   is  applied to the spring located between these masses ensuring that the strain in this spring remains the same all the time. The resulting system of equations can be written in the form:  
\begin{equation}
\ddot{u}_j(t)=\sigma(u_{j+1}-u_{j})-\sigma(u_{j}-u_{j-1})-A(\delta_{jj_0}+\delta_{j(j_0+1)})\sin(\nu t),
\end{equation}
where $\delta_{ij}$ is  the Kronecker delta, $\nu=k_1V$ and $k_1$ is the first real positive root of the equation $L(k)=0$. 
%


We performed simulations with   100 transition events   taking place before the waves reflected from the boundaries of the chain took any effect. The defect was shown to  approach   the steady-state TW regime for several values of  velocity $V$ which suggests that the corresponding analytical solution is a dynamic attractor.  
 
\begin{figure}[!htbp]
\includegraphics[scale=1]{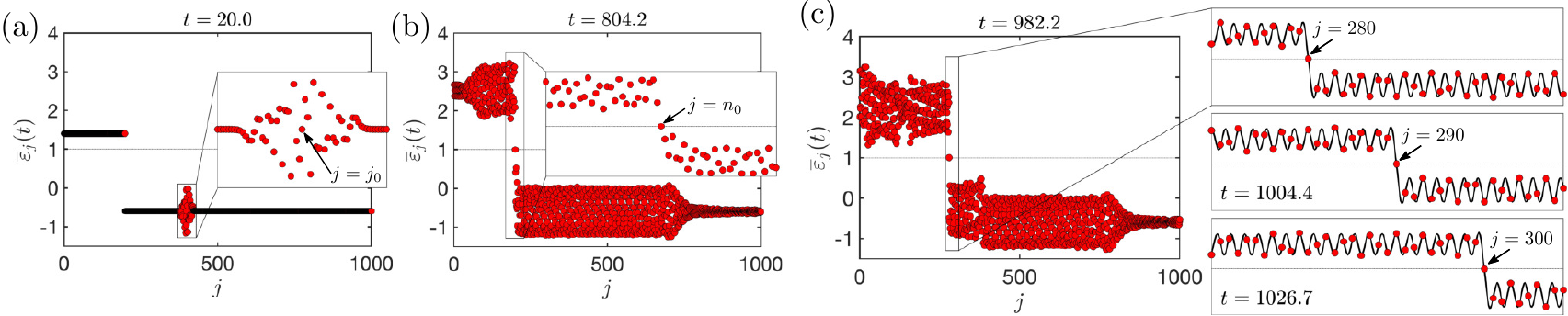}
\caption{Snapshots of the normalized strains $\overline{\varepsilon}_j(t)=\varepsilon_j(t)/\varepsilon_c$ at different moments: (a) initial propagation of the wave from the AC forces centred at $j=j_0=400$ and $j_0+1$; the initial defect is at $j=n_0=200$; (b) the first transition event takes place at $j=n_0$ soon after the energy from the DC force (turned on at $t=t_0=600$) at the left end arrived; (c) steady-state propagation achieved when the comparison with the analytical solution (black solid lines) is possible; the insets on the right show snapshots of strains when the front is at $j=$280, 290 and 300. 
}
\label{fig:Numerics}
\end{figure}

 
Our  Fig.~\ref{fig:Numerics} shows time evolution of the strain field.  The insets in Fig.~\ref{fig:Numerics}(c) show comparison between the numerically obtained data and the  analytical solution  for  $A=0.6,V=0.45$ and $\varepsilon_-=2.21$. In this case  $\varepsilon_+=-0.29$ and $G^M+G^m=0.27$ (the other parameters are $\sigma_0=2,\,\varepsilon_c=1$).
 In the attached Movie 1 we show the dynamic propagation of the phase boundary as demonstrated in the insets of Fig.~\ref{fig:Numerics}(c). The numerical solution is interposed with the analytical solution of the front moving at $V=0.45$. During the  propagation of the defect, the points progressively move from the phase with $\varepsilon_+$ to the phase with $\varepsilon_-$ by passing $\varepsilon_c$. The trajectories   follow precisely the analytical solution.

\begin{figure}[!htbp]
\includegraphics[scale=1]{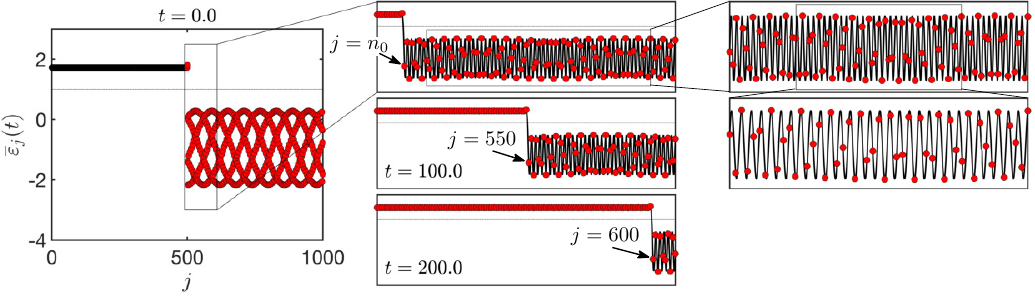}
\caption{
Normalized strains $\overline{\varepsilon}_j(t)=\varepsilon_j(t)/\varepsilon_c$ at different moments  with the initial data corresponding to the TW solution \eqref{eq:Solution} with the initial front position at  $j=n_0=500$ and propagation speed $V=0.5$. The inserts show comparison with the analytical results (black solid lines). 
}
\label{fig:Numerics_A_star}
\end{figure}

To show  stability of the frictionless  solutions we performed  numerical simulations with  initial conditions corresponding to the analytical solution \eqref{eq:Solution}. The time dependent  problem $\ddot{\varepsilon}_j=\sigma(\varepsilon_{j+1})+\sigma(\varepsilon_{j-1})-2\sigma(\varepsilon_{j})$ was  solved  while the ends of the chain were let  free: $\ddot{\varepsilon}_1=\sigma(\varepsilon_{2})-2\sigma(\varepsilon_{1})$ and $\ddot{\varepsilon}_N=\sigma(\varepsilon_{N-1})-2\sigma(\varepsilon_{N})$ with $N=1000$. The initial position of the defect was set  in the middle  of the chain at $n_0=500$. The snapshots of the moving front for this case are  shown in Fig.~\ref{fig:Numerics_A_star} where the propagation speed is   $V=0.5$. The attached Movie 2 demonstrates transient propagation of the  defect shown in Fig.~\ref{fig:Numerics_A_star}. We see that the analytically predicted  velocity $V$ is maintained  and the time dependence of strains follow   the analytical solution.

\section{FK problem: analytical results}

By performing integration in Eq. 7 in the main text we obtain the expression for the displacement field 
\begin{equation}
u(\eta)=\begin{cases}
  u_+ + \sum_{j=1}^K A_j\sin{(k_j\eta + \varphi_j)} -  \sum\limits_{k_j \in Z^+}\dfrac{\sigma_0}{k_j L'(k_j)}e^{-ik_j\eta},\, \eta>0,\\
 u_- + \sum_{j=1}^K A_j\sin{(k_j\eta + \varphi_j)} +\sum\limits_{k_j \in Z^-}\dfrac{\sigma_0}{k_j L'(k_j)}e^{-ik_j\eta},\, \eta<0.
\end{cases}
\label{eq:Solution_FK}
\end{equation}
 The sets $Z^\pm$ are defined in the   same way as in the FPU problem.  
 


The switching condition  $u(0)=u_c$ can be again written in two equivalent forms 
$
u_\pm=u_c\mp(\sigma_0/2)+\sigma_0R-\sum_jA_j\sin\varphi_j,
$
where 
\begin{equation}
R=\frac{1}{2}\sum\limits_{k_j \in Z_r^+}\frac{1}{k_j L'(k_j)}-\frac{1}{2}\sum\limits_{k_j \in Z_r^-}\frac{1}{k_j L'(k_j)}.
\end{equation} 



Next we compute the rate of energy dissipation ${\cal R}_\pm=G_\pm V$  following the same methodology as in the FPU problem.  We obtain:
\begin{equation}
G_\pm=\sum_{k_j\in Z_r^\pm,k_j>0}\left[\left(\pm 2\frac{\sigma_0}{k_jL'(k_j)}-A_j\sin\varphi_j\right)^2+A_j^2\cos^2{\varphi_j}\right]\omega^2(k_j)\left|\frac{\omega'(k_j)}{V}-1\right|.
\label{eq:EnergyFluxes_dislocation}
\end{equation}
The energy balance \eqref{eq:EnergyBalance} remains  the same and can be again verified by direct substitution. 
 From \eqref{eq:EnergyFluxes_dislocation} one can see that  the choice $\varphi_j=\varphi_j^*=\pi/2$ and $A_j=A_j^*=(-1)^j\sigma_0V/(\omega(k_j)^2(\omega'(k_j)-V)),\,j=1,2,...,K$ ensures frictionless propagation of the defect. 

\section{PB problem: analytical results}

With the TW ansatz applied, the Fourier transform reduces Eq. 11 to 
\begin{equation}
L(k)\hat{u}^+(k)+\hat{u}^-(k)=\frac{\hat{q}(k)}{\omega_-^2(k)-(Vk)^2},
\label{eq:Crack_Fourier}
\end{equation}
where superscripts $\pm$ define complex-valued functions which are analytic in the half-planes $\pm \text{Im }k>0$, respectively. To represent the external DC/AC  driving on the  boundary  of the chain,   the function $\hat{q}(k)$ must be chosen to have a zero  physical space image $q(\eta) \equiv0$. The kernel function $L(k)=(\omega_+^2(k)-(Vk)^2)/(\omega_-^2(k)-(Vk)^2)$ has  zeros $z_j$ (roots of $\omega_+^2(z_j)=(Vz_j)^2$)  and poles $p_j$  (roots of $\omega_-^2(p_j)=(Vp_j)^2$). The symmetry properties $L(-k)=L(k)$ and $L(\overline{k})=\overline{L(k)}$ remain here the same as in FPU and FK problems. We can then define the sets of poles  $P^\pm=P_c^\mp \cup P_r^\pm$ such that 
$P^\pm_c=\left\{p:\,\omega_+^2(p)-(pV)^2=0,\, \pm \text{Im }p>0\right\}$ and 
$P^\pm_r=\left\{p>0\,:\,\omega_+^2(p)-(pV)^2=0,\,\text{Im }p=0,\, \pm (\omega_-'(p)-V)>0\right\}$. Similarly, we define the sets of zeros: $Z^\pm=Z_c^\mp \cup Z_r^\pm $ with 
$Z^\pm_c=\left\{z:\,\omega_+^2(z)-(zV)^2=0,\, \pm \text{Im }z>0\right\}$ and 
$Z^\pm_r=\left\{z>0\,:\,\omega_+^2(z)-(zV)^2=0,\,\text{Im }z=0,\, \pm (\omega_+'(z)-V)>0\right\}$.


The problem \eqref{eq:Crack_Fourier} can be solved using  the Wiener-Hopf technique~\cite{noble1958methods}. The main step is the  factorization of the function $L(k)=L^-(k)/L^+(k)$. The standard factorization formula gives on  the real line 
\begin{equation}
L^\pm(k)=L^{\mp 1/2}(k)\exp{\left(-\frac{1}{2\pi i}\text{ p.v.} \int_{-\infty}^{\infty}\frac{\text{Log }L(\xi)}{\xi-k}\,d\xi\right)}
\label{eq:Factorisation_integral}
\end{equation}
We can alternatively  apply the Weierstrass factorization theorem and present the factors as infinite products~\cite{noble1958methods}. Then we obtain the representation 
\begin{equation}
L^\pm(k)=\left(\frac{\gamma}{1-V^2}\right)^{\mp1/2}(0\mp ik)^{\pm 1}\left[\frac{\prod_{z_j\in Z_r^\pm}\left(1-(k/z_j)^2\right)}{\prod_{p_j\in P_r^\pm}\left(1-(k/p_j)^2\right)}\frac{\prod_{z_j\in Z_c^\pm}\left(1-(k/z_j)\right)}{\prod_{p_j\in P_c^\pm}\left(1-(k/p_j)\right)}\right]^{\mp 1}\frac{1}{S},
\end{equation}
where for future convenience  we defined
\begin{equation}
S
=\frac{\prod_{z_j\in Z_r^+} z_j}{\prod_{p_j\in P_r^-} p_j}\frac{\prod_{z_j\in Z_r^-} z_j}{\prod_{p_j\in P_r^+} p_j}.
\end{equation}
These expressions can be evaluated if we know  the location of the zeros $z_j$ and the poles $p_j$ introduced above.

We can now rewrite the left hand side of \eqref{eq:Crack_Fourier}  as a sum of "+" and "-" functions that are  analytic in the upper and lower half plane, respectively:
\begin{equation}
\frac{1}{L^+(k)}\hat{u}^+(k)+\frac{1}{L^-(k)}\hat{u}^-(k)=\Psi(k).
\label{eq:Crack_WH}
\end{equation}
The explicit solution of \eqref{eq:Crack_WH} can be now obtained by decomposing the right hand side 
\begin{equation}
\Psi(k)=\frac{1}{L^-(k)}\frac{\hat{q}(k)}{\omega_-^2(k)-(Vk)^2}
\end{equation}
into  a sum of "+" and "-" functions. 
To represent the general DC and AC  sources we can  set 
\begin{equation}
\Psi(k)=2\pi C\delta(k)+2\pi \sum_{k_j} C_j\delta(k-k_j),
\end{equation}
where the wave numbers $k_j$ are chosen from  the set  $Z_r^- $ if the AC source is located ahead of the defect  and from the set  $P_r^+$ if the source is behind the defect. Since $\quad L^-(k_j)\left[\omega_-^2(k_j)-(Vk_j)^2\right]=0$, we have $q(\eta)=0$ and the sources,  parametrized by the constants $C$ (DC driving) and $C_j$ (AC driving),  are indeed invisible in the bulk.
%
%
%

If we further additively factorize the delta functions  $\delta(k-k_j)=2\pi[1/(0+i(k-k_j))+1/(0-i(k+k_j))]$ we can write 
\begin{equation}
\Psi(k)=\Psi^+(k)+\Psi^-(k),\quad \Psi^\pm(k)=\frac{C}{0\mp ik}+\sum_{k_j\in Z_r^-\cup P_r^+} \frac{A_j}{2}\left[\frac{e^{-i(\varphi_j-\pi/2)})}{0\mp i(k-k_j)}+\frac{e^{i(\varphi_j-\pi/2)}}{0\mp i(k+k_j)}\right].
\end{equation}
In this representation the complex amplitudes  $C_j$  are replace by the real  amplitudes $A_j$ and real phases $\varphi_j$. The total number of the corresponding  sinusoidal waves  is $K=|Z_r^-|+|P_r^+|$.

%
%

 We can now apply the Liouville theorem \cite{noble1958methods} to  \eqref{eq:Crack_WH}   and   obtain  the  explicit solution of our problem 
$
 \hat{u}^\pm(k)=L^{\pm}(k)\Psi^\pm(k).
$
Given that  $\hat{u}(k)=\hat{u}^+(k)+\hat{u}^-(k)$ we obtain in the physical space
$
u(\eta)=
\frac{1}{2\pi}\int_{-\infty}^{\infty}L^\pm(k)\Psi^{\pm}(k)e^{-ik\eta}\,dk,
$ when  $\pm\eta>0$, respectively, 
which gives Eq. 12 and Eq. 13 in the main text.
We can now apply  the switching condition to obtain 
$C+\sum_{j=1}^K A_j\sin{\varphi_j}=u_c$. Then 
using the boundary condition at $-\infty$ we obtain the  link  between the constant $C$ and the amplitude of the DC driving $\tau$  in the form  $\tau=(C/S(V))\sqrt{(1-V^2)/\gamma}$. In the physical space the ensuing solution   takes a form $u^{\pm}(\eta)=u_{1}^{\pm}(\eta)+u_{2}^{\pm}(\eta)$,  for  $\eta$ larger (smaller) than $0$, respectively.  Here 
\begin{equation}
u_{1}^{\pm}(\eta)=\sum_{z_j\in Z_r^+/P_r^-} \alpha_j^{\pm}\cos{(z_j \eta + \beta_j^{\pm})}+\sum_{z_j\in Z_c^+/P_c^-} \alpha_j^{\pm}e^{-i(z_j\eta+\beta_j^{\pm})}
\end{equation}
are the terms which do not contain DC/AC driving amplitudes explicitly. However, in contrast to the previous cases, the implicit dependence is present through the real coefficients $\alpha_j^\pm$ and $\beta_j^\pm$  representing   the  complex numbers:
\begin{equation}
\alpha_j^{\pm}e^{-i\beta_j^{\pm}}=\frac{i\Psi^{\pm}(k_j)L^{\mp}(k_j)\gamma}{2k_jV(\omega_{\pm}'(k_j)-V)},
\end{equation}
where $k_j=z_j$ when the sign is $+$ and $k_j=p_j$ if it is $-$.
The  part of the solution explicitly   related to external driving can be in turn  split into a DC and an AC  related  parts:   $u_{2}^{\pm}(\eta)=u_{DC}^{\pm}(\eta)+u_{AC}^{\pm}(\eta)$. Here $u_{DC}^{+}(\eta)=0$ and 
\begin{equation}
u_{DC}^{-}(\eta)= \frac{C}{S}\sqrt{\frac{1-V^2}{\gamma}}\left[\left(\sum_{z_j\in Z_c^-}\frac{i}{z_j}-\sum_{p_j\in P_c^-}\frac{i}{p_j}\right)-\eta\right].
\end{equation}
The AC related term is 
\begin{equation}
u_{AC}^{\pm}(\eta)= \sum_{k_j\in Z_r^-/P_r^+}A_j|L^{\pm}(k_j)|\sin{(k_j\eta+\varphi_j-\text{arg }L^{\pm}(k_j))}. 
\end{equation}

 In the main text we showed that in this problem the macro-level energy release rate is $V G^M(V)=(\tau^2(1-V^2)/2-(\gamma u_c^2)/2)V$ while   the rate of dissipation due to the AC driving is:
\begin{equation}
V G^m(V)=\sum_{z_j\in Z_r^-} \frac{A_j^2|L^+(z_j)|^2}{2}\omega_+^2(z_j)(V-\omega'_+(z_j))+\sum_{p_j\in P_r^+} \frac{A_j^2|L^-(p_j)|^2}{2}\omega_-^2(p_j)(\omega'_-(p_j)-V).
\end{equation}
The  dissipation due to  radiated elastic waves is:
\begin{equation}
VG_+(V)=\sum_{z_j\in Z_r^+} \frac{(\alpha_j^+)^2}{2}\omega_+^2(z_j)(\omega'_+(z_j)-V),\quad VG_-(V)=\sum_{p_j\in P_r^-} \frac{(\alpha_j^-)^2}{2}\omega_-^2(z_j)(V-\omega'_-(p_j)).
\end{equation}
The validity of the energy balance \eqref{eq:EnergyBalance} in this case was  checked numerically for the whole range of velocity $0<V<1$.

The total  dissipation becomes equal to zero  if $\alpha_j^+=0$ and $\alpha_j^-=0$ which can be ensured  if we  adjust the amplitudes $A_j$ in such a way that   $\Psi^+(z_j)=0$ and $\Psi^+(p_j)=0$.  These conditions can be rewritten as  a  linear system for the amplitudes $A_j$ :
\begin{equation}
\sum_{k_j\in Z_r^-\cup P_r^+}A_j\frac{k_j^2}{k_j^2-p_{i}^2}=u_c,\quad p_i\in P_r^-,\qquad \sum_{k_j\in Z_r^-\cup P_r^+}A_j\frac{k_j^2}{k_j^2-z_{i}^2}=u_c,\quad z_i\in Z_r^+
\end{equation}
with additional requirement that $\varphi_j=\varphi_j^*=\pi/2$.
